\documentclass[conference]{IEEEtran}
\IEEEoverridecommandlockouts
\usepackage{cite}
\usepackage{amsmath,amssymb,amsfonts}
\usepackage{algorithmic}
\usepackage{graphicx}
\usepackage{textcomp}
\usepackage{xcolor}
\def\BibTeX{{\rm B\kern-.05em{\sc i\kern-.025em b}\kern-.08em
    T\kern-.1667em\lower.7ex\hbox{E}\kern-.125emX}}
\begin{document}

\title{Quantum-Enhanced Generative Adversarial Networks: Comparative Analysis of Classical and Hybrid Quantum–Classical Generative Adversarial Networks\\
}

\author{\IEEEauthorblockN{Kun Ming Goh}
\IEEEauthorblockA{\textit{School of Computing} \\
\textit{Singapore Polytechnic}\\
Singapore, Singapore \\
https://orcid.org/0009-0008-7666-781X}
}

\maketitle

\begin{abstract}
Generative adversarial networks (GANs) have emerged as a powerful paradigm for producing high-fidelity data samples, yet their performance is constrained by the quality of latent representations, typically sampled from classical noise distributions. This study investigates hybrid quantum–classical GANs (HQCGANs) in which a quantum generator, implemented via parameterised quantum circuits, produces latent vectors for a classical discriminator. We evaluate a classical GAN alongside three HQCGAN variants with 3, 5, and 7 qubits, using Qiskit’s AerSimulator with realistic noise models to emulate near-term quantum devices. The binary MNIST dataset (digits 0 and 1) is used to align with the low-dimensional latent spaces imposed by current quantum hardware. Models are trained for 150 epochs and assessed with Fréchet Inception Distance (FID) and Kernel Inception Distance (KID). Results show that while the classical GAN achieved the best scores, the 7-qubit HQCGAN produced competitive performance, narrowing the gap in later epochs, whereas the 3-qubit model exhibited earlier convergence limitations. Efficiency analysis indicates only moderate training time increases despite quantum sampling overhead. These findings validate the feasibility of noisy quantum circuits as latent priors in GAN architectures, highlighting their potential to enhance generative modelling within the constraints of the noisy intermediate-scale quantum (NISQ) era.
\end{abstract}


\section{Introduction}
As Generative AI (GenAI) becomes increasingly prevalent in modern society, GANs have emerged as a powerful method for producing realistic data from Latent Noise. Despite their success, classical GANs often face challenges such as Mode Collapse, Unstable Training Dynamics, and Limited Representational Capacity. With advancements in Quantum Computing, Quantum Circuits offer potential improvements to Classical GAN through Higher-dimensional Representations and Entanglement-based Encoding.

\subsection{Background}

GANs have become a cornerstone of generative modeling in machine learning, especially for image generation tasks. A GAN consists of two competing neural networks, the generator and the discriminator,, trained simultaneously in an adversarial fashion~\cite{b1}. The Generator role is to map random noise to synthetic data, while the Discriminator evaluates a given data sample and determines if the data sample is real (i.e. from training set) or fake (i.e. produced by Generator). Through the training process, the generator is optimised to 'fool' the Discriminator, while the Discriminator is optimized to classify real and fake data. Through this process, the generator will ideally generate realistic data.
However, the training process for GAN is notorious for being rigorous and prone to failure modes. One major issue is when the GAN is only capable of producing a small subset of the training data distribution, leading to negatively affected GAN diversity. Another major issue is unstable convergence, which occurs when the ‘delicate’ adversarial training leads to Oscillations or Divergence instead of a stable equilibrium (ie, the discriminator overpowers the generator too quickly, or vice versa), causing the training to fail.
Given these limitations of Classical GANs and with the advancement in Quantum Technologies, researchers are looking at Quantum Computing as a method to potentially advance Gen AI. Quantum Computing leverages Qubits capable of existing in superposition and entanglement, theoretically allowing for an exponentially large state space to represent probability distributions. Theoretically, a Quantum-based Generator can encode rich distributions in the amplitudes of a Quantum State leading to the process of information in a high-dimensional Hilbert space that grows as $2^n$ for $n$ Qubits. This will allow a quantum generator to theoretically capture more complex patterns and correlations compared to a Classical Generator~\cite{b2}.

\subsection{Motivation}

The motivation for exploring a hybrid quantum-classical GAN stems from the practical challenges encountered during GAN training, such as Mode Collapse and Unstable Training Dynamics, as well as the theoretical potential of Quantum Computing. During training of GANs, various classical GAN models exhibited these issues despite numerous mitigation efforts.

Quantum computing introduces new capabilities, such as data representation through Superposition and Entanglement, which opens up opportunities for generative modelling. These quantum properties make it timely to investigate whether a Quantum Generator can enhance the capabilities of classical GANs.

\subsection{Research Question and Objectives}
The central research question is:
\begin{quote}
\emph{Can a Hybrid Quantum–Classical GAN generate binary digit images with quality and diversity comparable to Classical GANs, and does the Quantum Generator improve training stability or mode coverage?}
\end{quote}

The objectives of this study are:
\begin{enumerate}
    \item Design and train a QGAN with a quantum generator and a classical discriminator.
    \item Evaluate and compare QGAN image generation quality, diversity, and training stability with classical GANs.
    \item Investigate whether quantum properties help mitigate mode collapse and unstable training.
\end{enumerate}

\section{Literature Review}\label{review}

\subsection{Classical GAN Research}

Despite Classical GANs showing tremendous success in Generative Modelling and Image Synthesis, they remain inherently difficult to train \cite{b3}. Many efforts to counteract these difficulties have led to architectural advances such as the Deep Convolutional GAN (DCGAN), which introduced convolutional layers within the architecture to enhance Data Feature Learning and improve Sample Quality ~\cite{b4}. To address Training Instability, Wasserstein GAN (WGAN) was introduced, replacing the standard loss with the Wasserstein distance for smoother gradients ~\cite{b5}. Originally from WGAN, WGAN with Gradient Penalty (WGAN-GP) was introduced to further improve Training Stability through the use of Gradient Penalty techniques ~\cite{b6}.

\subsection{Quantum GAN Research}

With advancements and breakthroughs in the field of Quantum Computing, different methods to Model Data through properties like Superposition and Entanglement have become increasingly available. Early efforts have shown that parametrised quantum circuits can be used in adversarial training loops, establishing the feasibility of QGANs ~\cite{b2}. A hybrid of Classical-Quantum GAN has also been theorised, which comprises a Quantum Generator and a Classical Discriminator. Despite the major claims and potential, the studies conducted are extremely limited to synthetic and low-dimensional data with no comparison within the lens of practical contexts.

\subsection{Quantum-Hybrid GAN Research}

Hybrid Quantum-Classical GANs (HQCGAN) have emerged as a new approach that integrates Quantum Computing with Classical Computing. A HQCGAN utilises a Quantum Circuit-Based Component for generative modelling with a classical deep learning neural network as the discriminator. In theory, this will create a highly accurate and efficient HQCGAN which surpasses the capabilities of Classical GAN.

The key advantages of HQCGAN is based on its ability to construct a latent distribution that is inherently more expressive due to the nature of quantum circuits. This equips the generator to capture complex data distribution in fewer dimensions, leading to efficient training and solving issues such as mode collapse and insufficient high-dimensional sampling without sacrificing diversity and representational power.

Quantum Circuits introduces entanglement natively (Qubits are not independent unlike Classical Bits). This factor implies that some aspects of feature correlation are already embedded in the latent space, leading to faster convergence with fewer data with better results ~\cite{b7}, achieving a balance between efficiency and quality.

These factors combined with the capabilities of a Classical Discriminator amplifies a stable and efficient GAN architecture where Quantum Noise enhances diversity and classical networks enforce realism. This approach achieves a theoretical balance between Quantum and Classical Computing.

\subsection{Research Gaps}

Although QGANs have been proven to work theoretically, little empirical research has been conducted to compare HQCGANs with Classical GANs in standard image generation operations. This raises various key questions that remain unanswered. Do Quantum Generators offer improved Mode Coverage? Is the training more stable? Is the risk of mode collapse reduced?

This paper will address these gaps identified through the training and evaluation of both HQCGAN and classical GAN. The purpose of this paper is to provide practical insight into the utility and limitations of HQCGANs in image synthesis.

\section{Experimental Results and Analysis}\label{Experimental Results and Analysis}

\subsection{Overview of Experimental Setup}

\subsubsection{Dataset: Binary MNIST}

The experiment setup utilises a filtered version of the Modified National Institute of Standards and Technology (MNIST) dataset, retaining only samples labelled as digits '0' and '1'. This binarisation process was intentionally designed to simplify the data distribution while preserving sufficient visual complexity for the evaluation of generative performance of GAN models. To address the class imbalance, undersampling was performed for both classes, where samples were randomly selected from the majority class until data counts for both classes matched. This is done to avoid mode collapse and ensure fairness of the experiment. Each image is of dimension $28 \times 28$ pixels and reshaped into a 784-dimensional vector prior to GAN training. All pixel values were also normalised to the range $[-1, 1]$ using the following formula:
\begin{equation}
x_{\text{norm}} = \frac{x - \min}{\max - \min} \times 2 - 1
\end{equation}

Normalisation ensures compatibility with the tanh output activation of the generator. The resulting dataset was then shuffled and batched with a batch size of 64. The usage of the MNIST dataset allows for clearer evaluation of generative performance under noise in quantum circuits, and ensures comparative studies between classical and HQCGAN remains interpretable and computationally tractable.

\subsubsection{Model Architectures}
The classical GAN implemented in this experiment adopts a fully connected feedforward architecture for both the generator and the discriminator, ideal for flattened binary MNIST dataset of shape $28 \times 28 = 784$. The classical GAN will serve as a basis for the evaluation (baseline) of the HQCGAN in this experiment.

The generator takes a latent input vector \( z \in \mathbb{R}^{100} \), sampled from a standard normal distribution. Subsequently, it passes through two hidden layers with ReLU activation, followed by a final layer with tanh activation to produce an output image vector \( \hat{x} \in \mathbb{R}^{784} \) in the range $[-1, 1]$. The forward pass formula is defined as follows:
\begin{equation}
G(z) = \tanh(W_3 \cdot \text{ReLU}(W_2 \cdot \text{ReLU}(W_1 \cdot z + b_1) + b_2) + b_3)
\end{equation}

The discriminator receives real or generated images and outputs a probability score. It consists of two hidden layers with LeakyReLU activation and a final sigmoid-activated output.

The GAN is trained using the standard non-saturating loss formulation with Binary Cross-Entropy (BCE) as the objective function. The discriminator loss formula is defined as follows:

\begin{equation}
\mathcal{L}_D = -\frac{1}{2} \left[\log D(x) + \log \left(1 - D(G(z))\right)\right]
\end{equation}

The generator loss formula is defined as follows:

\begin{equation}
\mathcal{L}_G = -\log D(G(z))
\end{equation}

These formulas are consistent across all GANs throughout the experiment, with architectural modifications being the only applied changes made for the HQCGAN.

\subsubsection{GAN Variants}
To evaluate the theoretical promises of HQCGAN, we implemented an HQCGAN and a Classical GAN to compare their performance. A summary of the configuration of each model is provided below:

\begin{table}[htbp]
\caption{HQCGAN and Classical GAN Model Configurations}
\begin{center}
\begin{tabular}{|c|c|c|c|c|}
\hline
\textbf{Model} & \textbf{Gen. Type} & \textbf{Disc. Type} & \textbf{Lat. Dim.} & \textbf{Qubits Used} \\
\hline
Classical  & Classical & Classical & 100 & 0 \\
HQCGAN-3   & Quantum   & Classical & 3   & 3 \\
HQCGAN-5   & Quantum   & Classical & 5   & 5 \\
HQCGAN-7   & Quantum   & Classical & 7   & 7 \\
\hline
\multicolumn{5}{l}{\footnotesize{Lat. Dim. = Latent Dimension.}}
\end{tabular}
\label{tab:HQCGAN_configs}
\end{center}
\end{table}

\subsubsection{Quantum Latent Sampling with Noise}

In HQCGAN, the latent vector ($z$) is not sampled from a traditional Gaussian distribution. Instead, it is derived from measurements of noisy quantum circuits simulated using the Qiskit AerSimulator. Each circuit consists of Hadamard Gates applied to all qubits. The resulting bitstrings will be interpreted as binary vectors and linearly scaled to the range $[-1, 1]$ using the following transformation:
\begin{equation}
z = 2 \cdot \text{bitstring\_sample} - 1
\end{equation}
To introduce realistic quantum effects, the circuits were executed on an AerSimulator backend with a custom noise model, incorporating the following: 
\begin{itemize}
\item Depolarising noise: simulates random Pauli errors on gates
\item Amplitude damping: emulates energy relaxation ($T_1$ decay)
\item Readout error: models inaccuracies in measurement outcomes
\end{itemize}
To validate the effect of quantum noise, the state density matrix $\rho$ was extracted and visualised using Qiskit tools, including:
\begin{itemize}
\item Cityscape plots to visualise real and imaginary components of $\rho$
\item Pauli vector representations to capture expectation values of Pauli observables
\item Bloch multivector diagrams for multi-qubit state structure
\end{itemize}

\subsubsection{Training Strategy and Metrics}\label{Training Strategy}
The training procedure followed the standard GAN loop, with both classical and HQCGAN optimised using the Adam optimiser. Key hyperparameters and metrics tracked throughout the training are detailed below:

\paragraph{Hyperparameters:}
The training of the GAN models was conducted using a batch size of 64 and run over 150 epochs to ensure sufficient exposure to the dataset. The optimiser selected was Adam, a widely used adaptive learning rate optimisation algorithm in deep learning. A learning rate of $0.0002$ was applied, which balances the need for stable convergence with the flexibility to escape local minima. This configuration reflects common best practices for training GANs on datasets of similar size and complexity.

\paragraph{Accuracy Metrics:}
To assess the quality and diversity of the generated images, two key accuracy metrics were used: the Fréchet Inception Distance (FID) and the Kernel Inception Distance (KID).

The FID score is defined mathematically as:
\begin{equation}
\text{FID} = \lVert \mu_r - \mu_g \rVert^2 + \text{Tr} \left( \Sigma_r + \Sigma_g - 2 (\Sigma_r \Sigma_g)^{\frac{1}{2}} \right)
\end{equation}
This metric evaluates the distance between the feature distributions of real and generated images, assuming both follow a multivariate Gaussian distribution. Lower FID values indicate greater similarity between the generated and real data distributions, signifying better visual fidelity and diversity.

The KID is formulated as:
\begin{equation}
\text{KID} = \left\lVert \frac{1}{m} \sum_{i=1}^{m} \phi(x_i) - \frac{1}{n} \sum_{j=1}^{n} \phi(y_j) \right\rVert^2
\end{equation}
Unlike FID, KID uses the Maximum Mean Discrepancy (MMD) with polynomial kernels and does not assume Gaussianity. This makes it particularly suitable for smaller datasets, offering an unbiased estimate of the similarity between real and generated feature representations.

\paragraph{Efficiency Metrics:}
In addition to accuracy, training efficiency was tracked using two practical metrics. The first is the \textbf{time per epoch}, which captures how long it takes the model to complete a single pass through the dataset; this is especially critical when comparing model architectures or performing hyperparameter tuning. The second is the \textbf{total number of samples seen} during training, which provides insight into the scale of data exposure and can help contextualise model convergence patterns and learning stability.

\subsection{Quantum Circuit Visualisation}
Quantum circuits were used to generate a latent vector ($z$) for the HQCGAN. These circuits were created with the help of the Qiskit AerSimulator, which supports realistic quantum noise modelling. This section presents a visual analysis of the quantum states produced by the circuits used to sample the latent space.

\subsubsection{Quantum Circuit Diagram}

\begin{figure}[htbp]
\centering
\includegraphics[width=0.75\linewidth]{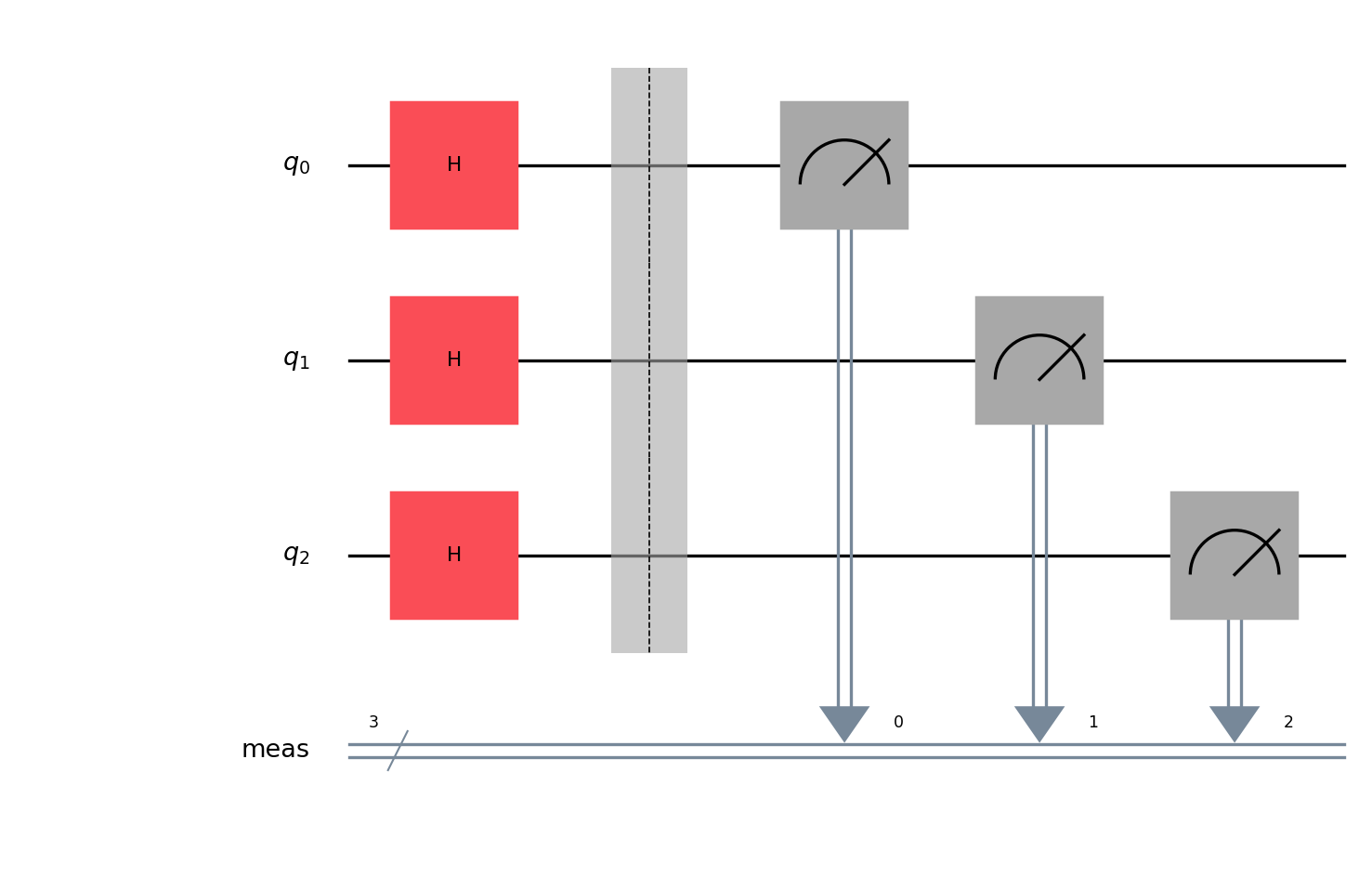}
\caption{Quantum circuit used for sampling latent vectors. Each qubit is initialised in $\left| 0 \right\rangle$ and transformed by Hadamard gates. Full measurement is performed post-noise.}
\label{fig:QuantumCircuit}
\end{figure}

The quantum circuit for each configuration ($n = 3, 5, 7$ qubits) consists of parallel Hadamard Gates, followed by a full measurement. Each qubit is initialised in the $\left| 0 \right\rangle$ state and transformed into a superposition state through the $H$ gate. This will ultimately provide a uniform sampling across all $2^n$ basis states. The quantum circuit is shown in \textbf{Fig.~\ref{fig:QuantumCircuit}}.

\subsubsection{Noisy Density Matrix (Cityscape)}
\begin{figure}[htbp]
\centering
\includegraphics[width=0.75\linewidth]{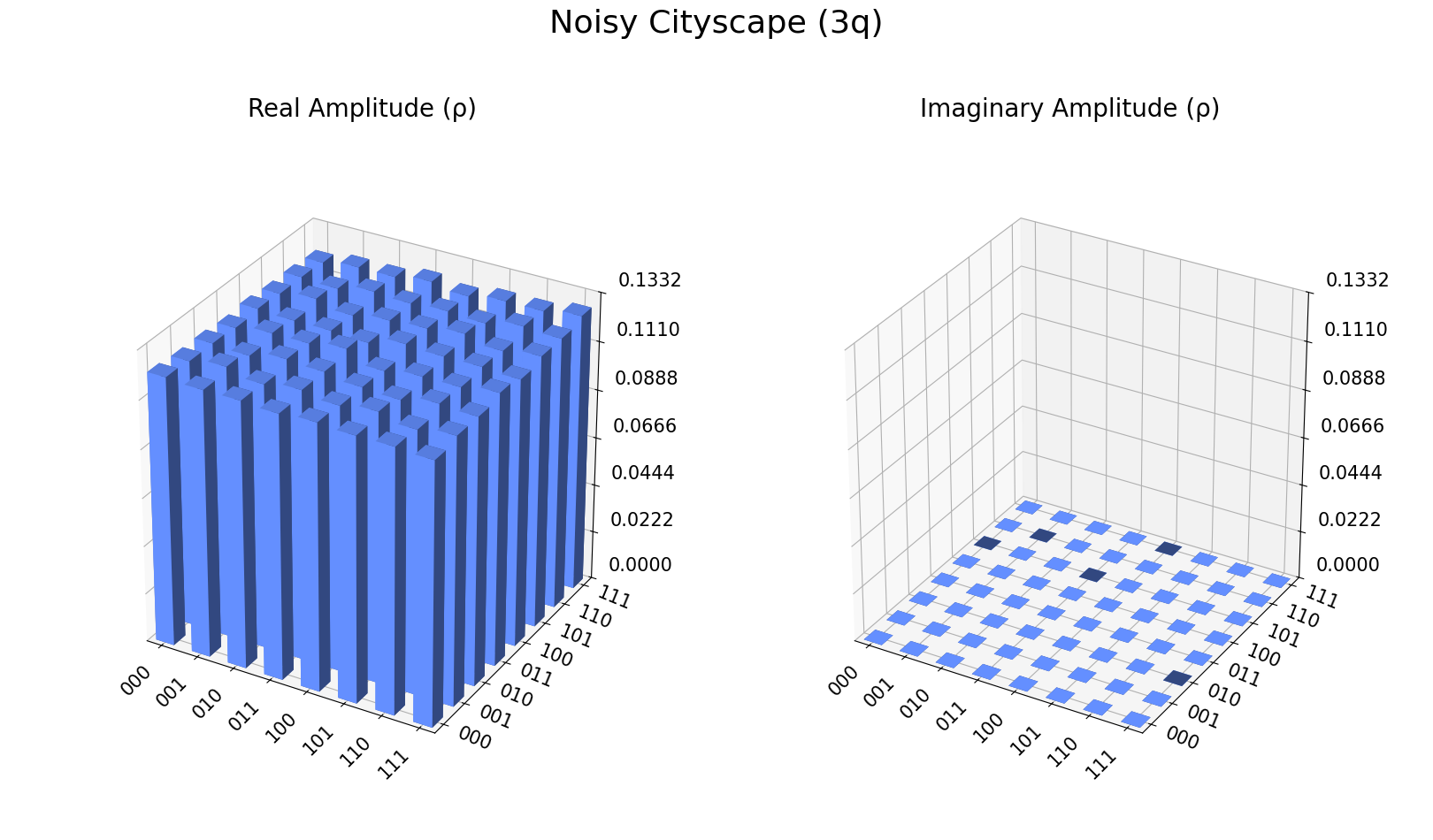}
\caption{Cityscape visualisation of the noisy quantum state density matrix. The z-axis represents the real component of the matrix elements after noise simulation.}
\label{fig:NoisyCityscape}
\end{figure}

To understand how quantum noise impacts the state, we visualise the real component of the noisy density matrix as a cityscape plot in \textbf{Fig.~\ref{fig:NoisyCityscape}}. This reveals how much of the quantum state remains coherent and decoherent due to the applied noise model.

\subsubsection{Pauli Vector Representation}
\begin{figure}[htbp]
\centering
\includegraphics[width=0.75\linewidth]{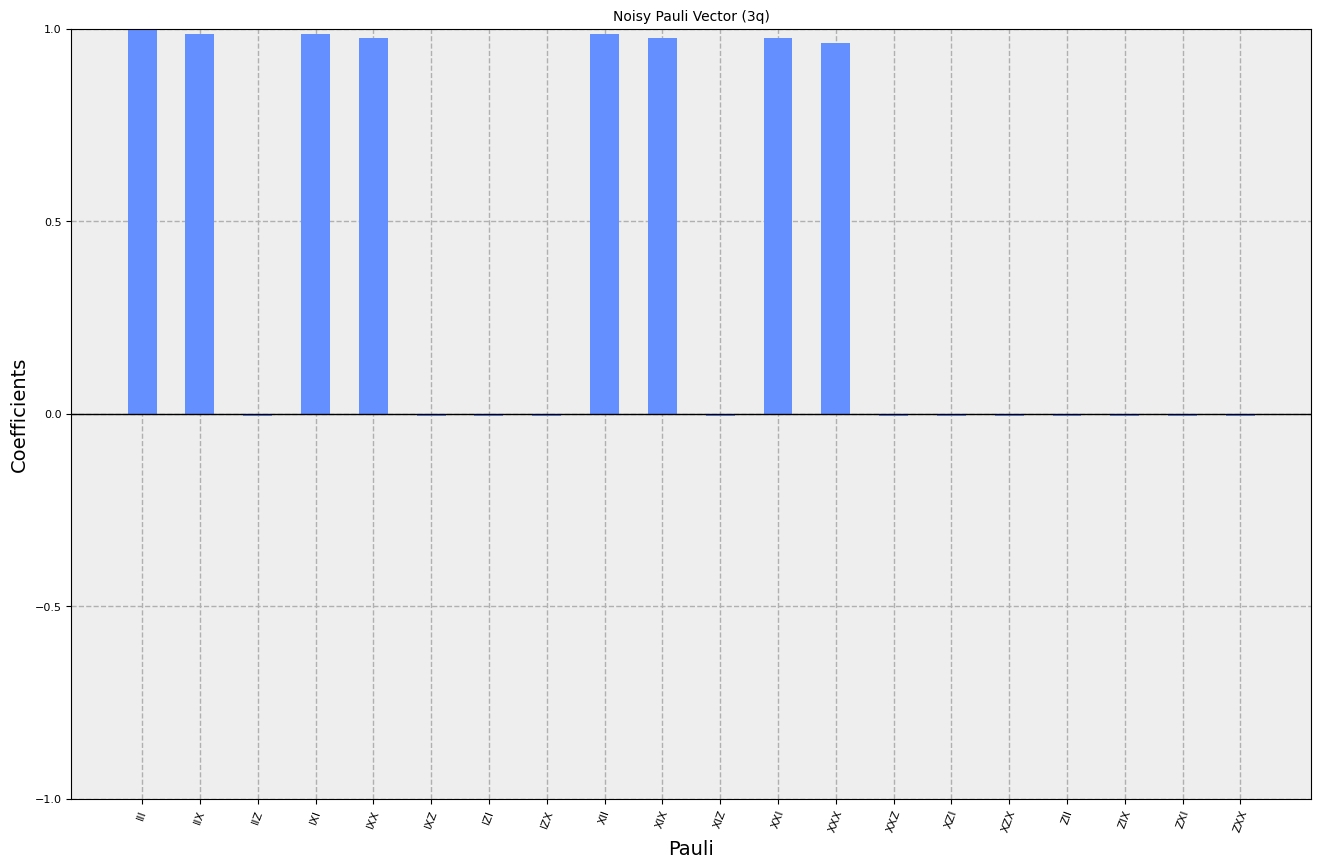}
\caption{Pauli vector representation of the noisy quantum state. Shorter vectors indicate loss of coherence due to amplitude damping and depolarising noise.}
\label{fig:PauliVector}
\end{figure}

The Pauli vector visualisation decomposes the quantum state into the basis of Pauli matrices ($I, X, Y, Z$). This is useful to evaluate the amount of remaining coherence and observable noise-induced shrinkage in the Bloch sphere as shown in \textbf{Fig.~\ref{fig:PauliVector}}.

\subsubsection{Bloch Multivector}
\begin{figure}[htbp]
\centering
\includegraphics[width=0.75\linewidth]{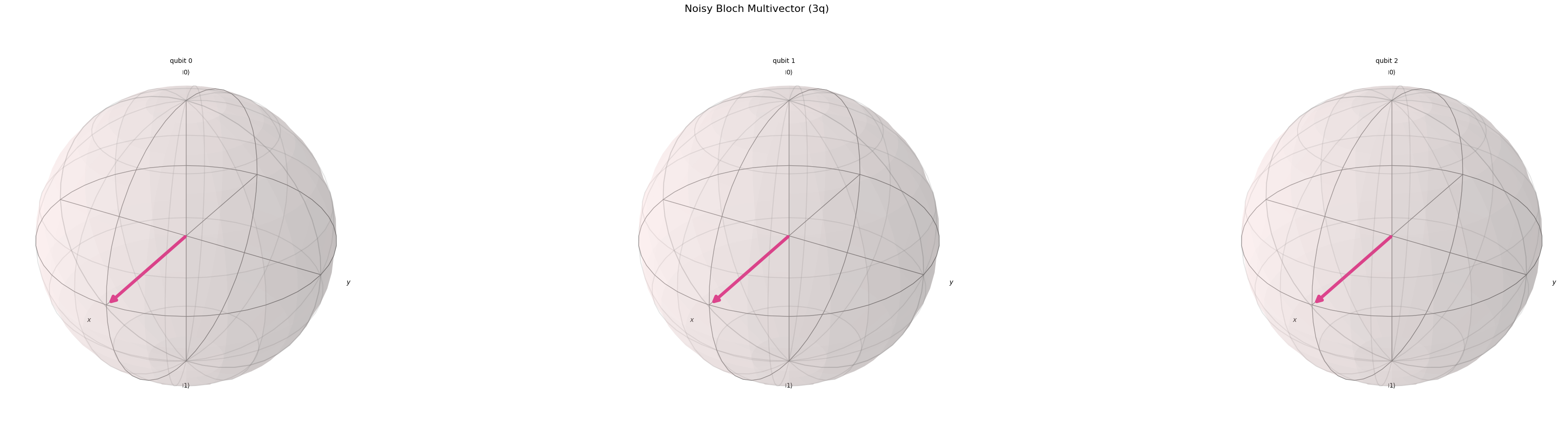}
\caption{Bloch multivector visualisation showing each qubit’s Bloch representation. Deviation from pure state vectors indicates decoherence and noise effects.}
\label{fig:BlochMultivector}
\end{figure}

The Bloch multivector plot represents the Bloch sphere components of each qubit individually. It helps visualise the influence of noise at the level of individual qubits, showing the deviation from the ideal pure state as shown in \textbf{Fig.~\ref{fig:BlochMultivector}}.

\subsubsection{Noisy Bitstring Count Histogram}
\begin{figure}[htbp]
\centering
\includegraphics[width=0.75\linewidth]{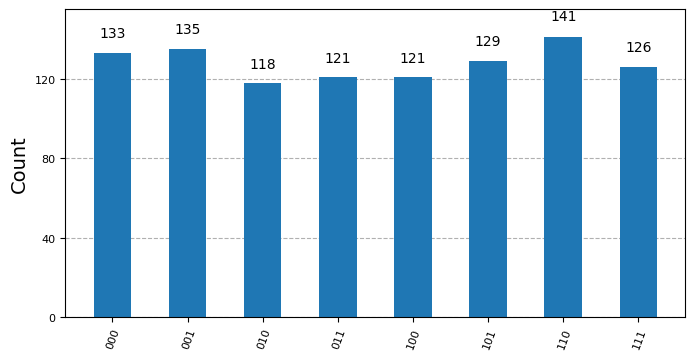}
\caption{Measurement outcome histogram from quantum circuit under noise simulation. Bitstrings represent sampled latent vectors, exhibiting minor deviations due to quantum noise.}
\label{fig:NoisyHistogram}
\end{figure}

The resulting measurement distribution is visualised in \textbf{Fig.~\ref{fig:NoisyHistogram}}. Ideally, in a noise-free simulation, each bitstring would occur with roughly equal frequency. However, the inclusion of depolarising and amplitude damping noise models introduces subtle biases into the sampling distribution. These imperfections reflect real-world quantum hardware limitations, and their impact is critical to evaluating the robustness of quantum-based latent encodings.

\subsection{Generator and Discriminator Training Dynamics}

\subsubsection{Discriminator Loss Dynamics}\label{Disc-Loss-Dyn}

The loss of the discriminator ($\mathcal{L}_D$) quantifies the degree of competence of the discriminator to distinguish between real and generated samples. Lower values indicate better discrimination performance, but excessively low loss may signal that the generator is underperforming or that the discriminator is overpowering the generator, potentially causing Mode Collapse.

\begin{figure}[htbp]
\centering
\includegraphics[width=0.75\linewidth]{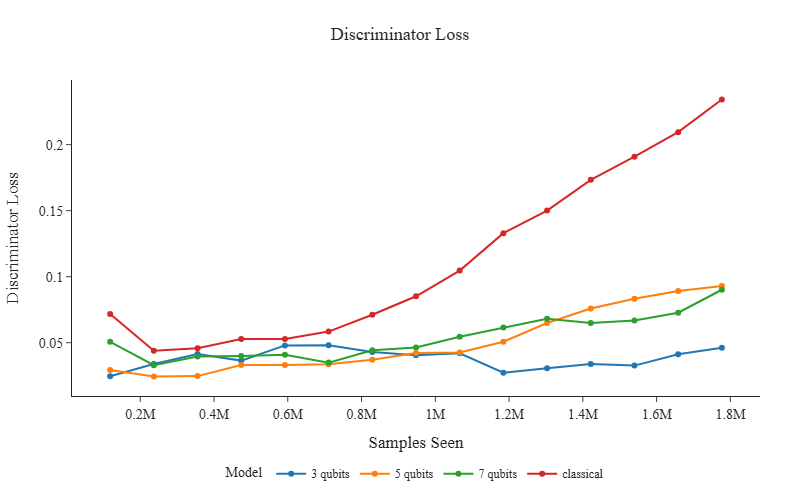}
\caption{Discriminator loss trends ($\mathcal{L}_D$) across training for classical GAN and HQCGANs with 3, 5, and 7 qubits.}
\label{fig:DiscLoss}
\end{figure}

As illustrated in \textbf{Fig.~\ref{fig:DiscLoss}}, the Classical GAN exhibits a sharp and consistent increase in $\mathcal{L}_D$ as the number of samples seen increases from around 0.7M. This observation suggests a failure to maintain equilibrium with its generator. This is a common problem with Classical GAN, where the generator fails to keep up with the discriminator, leading to poor-quality outputs, inflated discriminator confidence, and mode collapse.

In contrast, all HQCGAN (3, 5, 7 qubits) demonstrate relatively stable $\mathcal{L}_D$, indicating a more sustained equilibrium adversarial training process. In particular, the 3 qubits HQCGAN (Blue Line) maintain the lowest and most stable $\mathcal{L}_D$ amongst the other HQCGANs. This implies that the generator was able to consistently fool the discriminator in training. The smooth gradient in these quantum-based losses suggests that quantum sampling may introduce latent diversity that supports more effective generator updates. 

These trends and observations empirically validate some theoretical advantages of the HQCGANs mentioned in Section~\ref{review}. Specifically, the $\mathcal{L}_D$ trajectories of the HQCGAN reflect improved training stability and latent diversity compared to the Classical GAN.

\subsubsection{Generator Loss Dynamics}

The loss of the generator ($\mathcal{L}_G$) quantifies the degree of competence of the generator in generating samples similar to the training data. Lower values indicate better generator performance, but excessively low loss may signal that the discriminator is underperforming or that the generator is overpowering the discriminator, potentially causing mode collapse.

\begin{figure}[htbp]
\centering
\includegraphics[width=0.75\linewidth]{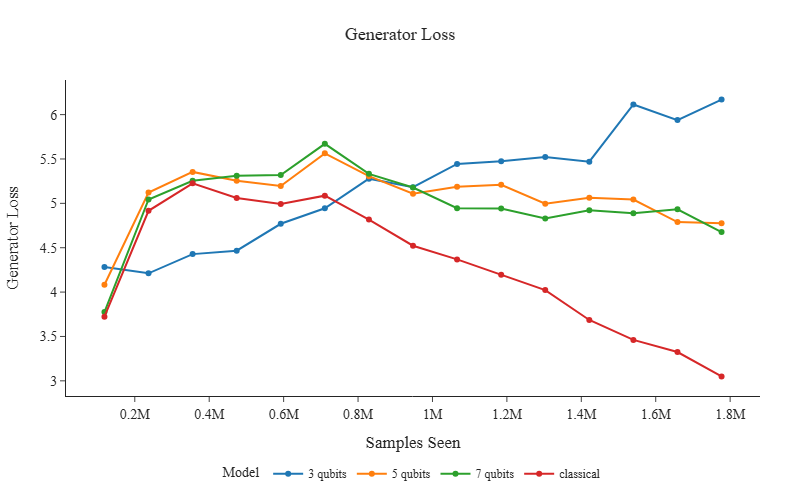}
\caption{Generator loss trends across training for classical GAN and HQCGANs with 3, 5, and 7 qubits.}
\label{fig:GeneratorLoss}
\end{figure}

As illustrated in \textbf{Fig.~\ref{fig:GeneratorLoss}}, the Classical GAN $\mathcal{L}_G$ steadily decreases when $N \approx 0.7M$, reaching values of $\mathcal{L}_G \approx 3$. Although this suggests that the generator is improving, the concurrent increase in $\mathcal{L}_D$ (from Section~\ref{Disc-Loss-Dyn}) suggests that this improvement may be artificial. The possibility of an overpowered generator dominating a weakening discriminator resulting in Mode Collapse cannot be ruled out in this scenario.

In contrast, all HQCGANs (3, 5, 7 qubits) exhibit stable $\mathcal{L}_G$ that oscillates within a narrow band where $4.5\leq\mathcal{L}_G\leq5.5$ throughout training. This oscillation is indicative of a healthy adversarial equilibrium. In particular, the 3-qubit HQCGAN display a slight upward gradient in $\mathcal{L}_G$ when $N \approx 1.4M$, potentially reflecting signs of under-fitting due to reduced latent dimensionality.

These dynamics observed reinforce the hypothesis that quantum-generated latent spaces introduce sufficient variability to prevent both premature convergence and over-fitting. The consistent $\mathcal{L}_G$ values suggest that quantum generators are constantly challenged by discriminators. Therefore, we maintain a productive adversarial training process in equilibrium without the occurrence of mode collapse.

\subsection{Combined Interpretation of Training Dynamics}

The adversarial equilibrium between the generator and the discriminator is critical for stable GAN training. The dual analysis of $\mathcal{L}_D$ and $\mathcal{L}_G$ in the classical GAN and HQCGAN reveals contrasting behavioural dynamics that underline the benefits of quantum-enhanced sampling.

In the classical GAN, the $\mathcal{L}_D$ increased significantly while $\mathcal{L}_G$ decreased steadily. This divergence suggests that the generator quickly exploited the fixed latent space and learnt to fool the discriminator too easily. Such a pattern is often seen as a symptom of mode collapse, where the generator output becomes overly repetitive, leading to poor diversity in generated samples despite a seemingly 'low' loss.

In contrast, the HQCGANs (3, 5, 7 qubits) demonstrated mutually stabilised loss profiles. $\mathcal{L}_D$ remained consistently low, indicating effective detection of generator outputs, while $\mathcal{L}_G$ maintained moderate values without sharp changes. This interplay implies a sustained adversarial challenge between the quantum generator and the classical discriminator, which encourages continuous improvements without mode collapse.

Moreover, the progressive increase in $\mathcal{L}_G$ with more qubits may reflect the growing complexity of the latent space. Higher qubit counts introduce richer and more entangled latent vectors, offering better diversity but also increasing the learning challenge for the generator.

In summary, the observed training dynamics suggests that HQCGAN not only solves the conventional problems faced by classical GANs, but also actively fulfils the theoretical advantages proposed in Section~\ref{review}. Specifically, it fulfils the theoretical advantage of enhanced expressiveness and diversity from quantum noise.

\subsection{FID and KID Evaluation}

To evaluate the realism and diversity of the generated samples, two established metrics were tracked throughout training: \textbf{Fréchet Inception Distance (FID)} and \textbf{Kernel Inception Distance (KID)}. The formulas and rationale for these metrics were introduced in Section~\ref{Training Strategy}.

\subsubsection{Fréchet Inception Distance (FID)}

\begin{figure}[htbp]
\centering
\includegraphics[width=0.75\linewidth]{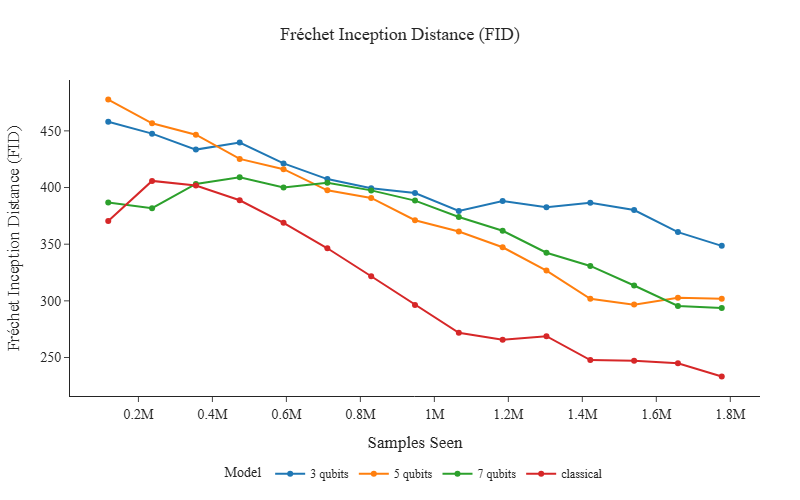}
\caption{Fréchet Inception Distance trends across training for classical GAN and HQCGANs with 3, 5, and 7 qubits.}
\label{fig:FID}
\end{figure}

\textbf{Fig.~\ref{fig:FID}} illustrates the progression of FID throughout the training of the classical GAN and HQCGAN. As expected, FID scores decreased consistently with the number of samples seen, indicating better alignment between the generated and true data distributions. Among the models, the classical GAN achieved the lowest overall FID scores, suggesting that it produced samples whose features most closely resembled those of the original dataset when evaluated using the Inception network.

Notably, 5-qubit and 7-qubit HQCGANs demonstrated competitive performance, particularly in the later stages of training. The 7-qubit model notably converged towards the classical GAN, indicating that greater quantum latent dimensionality may enhance sample fidelity. In contrast, the 3-qubit HQCGAN plateaued at a higher FID value where $N \approx 1$M, potentially due to the reduced representational capacity in the limited latent space dimensionality.

\subsubsection{Kernel Inception Distance (KID)}

\begin{figure}[htbp]
\centering
\includegraphics[width=0.75\linewidth]{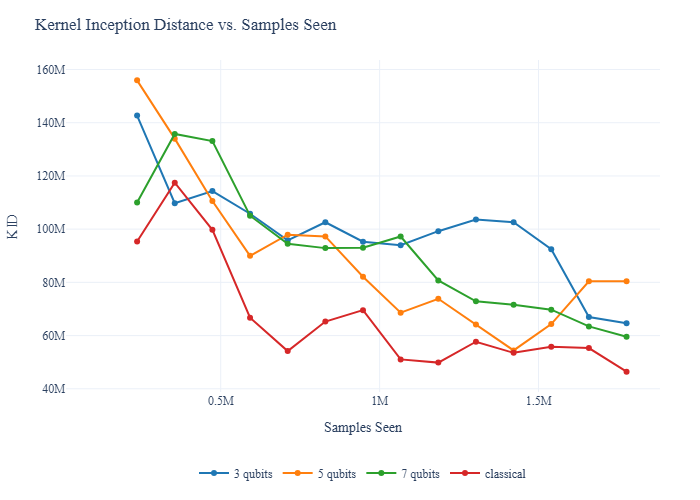}
\caption{Kernel Inception Distance trends across training for classical GAN and HQCGANs with 3, 5, and 7 qubits.}
\label{fig:KID}
\end{figure}

\textbf{Fig.~\ref{fig:KID}} illustrates the Kernel Inception Distance (KID) curves over the training duration for the classical GAN and HQCGAN. KID, unlike FID, is based on the Maximum Mean Discrepancy (MMD) without the assumption of Gaussianity in the distribution of Inception features. This makes KID a more robust and unbiased measure, especially when the sample size is limited or the generative model produces outputs with diverse characteristics.

Across all models, KID scores generally decreased as training progressed, indicating improved alignment with the real data distribution. Although the classical GAN initially exhibited higher KID scores where $N \approx 0.4$M, it quickly outperformed all the HQCGANs. However, the 7-qubit HQCGAN consistently achieved competitive KID scores, especially beyond $N \approx 1.1$M sample seen. This suggests that increasing quantum latent dimensionality improves not only feature diversity but also robustness in distributional learning provided sufficient qubit is involved.

Given its impartial nature, KID should be treated as the primary metric to compare the performance of the GANs in this study. It captures subtle discrepancies that FID may overlook, particularly in models where Gaussian assumptions do not hold.

\subsubsection{Evaluation Metrics Discussion}

\textbf{Table~\ref{tab:metrics_evaluation}} illustrates the final evaluation metrics between the classical GAN and the HQCGAN (3, 5, 7 qubits). The sample seen ($N$) is consistent throughout all the GAN models. The final numerical values are derived from the trend curve in \textbf{Fig.~\ref{fig:FID}} and \textbf{Fig.~\ref{fig:KID}}.

\begin{table}[htbp]
\caption{Metrics evaluation across models}
\begin{center}
\begin{tabular}{|l|c|c|c|c|}
\hline
\textbf{Model} & \textbf{Classical} & \textbf{HQCGAN-3} & \textbf{HQCGAN-5} & \textbf{HQCGAN-7} \\
\hline
Final FID        & 201   & 330   & 307   & 294   \\
Final KID    & 46.4M  & 64.6M  & 80.5M  & 59.5M  \\
\hline
\end{tabular}
\label{tab:metrics_evaluation}
\end{center}
\end{table}

With reference to Table~\ref{tab:metrics_evaluation}, the results further substantiate the hypothesis that increased quantum latent dimensionality improves the representational power of quantum generators. Although HQCGAN-3 struggled to match the classical GAN, most likely due to under-parameterisation, HQCGAN-7 closely matched the performance of the classical GAN’s performance, especially in the KID metric. The observed trends affirm the theoretical expectations of QGANs and HQCGANs, suggesting that quantum noise-injected latent sampling paired with classical discriminators can generate diverse and distribution-consistent data, provided that the quantum circuit has sufficient capacity.

\subsection{Efficiency and Scalability}

In addition to evaluating generative capabilities through quantitative metrics such as FID and KID, it is equally crucial to evaluate the efficiency and scalability of HQCGANs due to the theoretical promises of more efficient training duration. These aspects become especially pertinent in the context of quantum-based models, where access to quantum hardware is limited, and simulation time on classical back-ends can be substantial.

\begin{table}[htbp]
\caption{Efficiency comparison across models}
\begin{center}
\begin{tabular}{|l|c|c|c|c|}
\hline
\textbf{Model} & \textbf{Cls.} & \textbf{H3} & \textbf{H5} & \textbf{H7} \\
\hline
Qubits & 0 & 3 & 5 & 7 \\
Lat. dim & 100 & 3 & 5 & 7 \\
Time/1M (s) & 186.51 & 203.71 & 203.98 & 212.00 \\
Q. shots & 0 & 23808 & 23808 & 23808 \\
\hline
\multicolumn{5}{l}{\footnotesize{Cls. = Classical;}}\\
\multicolumn{5}{l}{\footnotesize{H3 = HQCGAN-3; H5 = HQCGAN-5; H7 = HQCGAN-7;}} \\
\multicolumn{5}{l}{\footnotesize{Lat. dim = latent dimension; Q. shots = quantum shots;}} \\
\multicolumn{5}{l}{\footnotesize{1M = one million.}}
\end{tabular}
\label{tab:efficiency_comparison}
\end{center}
\end{table}

\textbf{Table~\ref{tab:efficiency_comparison}} presents the efficiency metrics across all GAN models. As expected, the classical GAN exhibits the lowest average run-time per million samples, due to its purely neural nature without any quantum simulation overhead. However, all HQCGAN models maintain comparable runtimes, with only a slight increase observed as the number of qubits rises from 3 to 7. This suggests that quantum latent sampling scales efficiently in simulation and may remain feasible on near-term quantum hardware.

Although classical GANs are faster, HQCGANs demonstrate respectable training efficiency and show promise in representing complex distributions with significantly reduced latent dimensionality. This highlights their potential as compressed generative architectures with competitive performance-to-cost trade-offs. The efficiency may also deviate from the theoretical lens due to the lack of quantum hardware and usage of quantum simulation.

\subsection{Summary of Findings}

This sub-section explored the implementation and comparative evaluation of classical GANs and HQCGANs on a binary variant of the MNIST dataset. Three HQCGAN configurations were tested, with 3, 5, and 7 qubits serving as quantum latent dimensions. A consistent generator and discriminator architecture was used across all models, with quantum circuits generating latent vectors subject to simulated device-level noise.

Visual inspection of the quantum circuit output confirmed high-entropy, non-degenerate sampling across all qubit variants. Despite reduced latent dimensionality, the HQCGANs—particularly the 7-qubit model—demonstrated competitive training dynamics and generative capability. The discriminator and generator loss curves showed convergence behaviours similar to the classical GAN.

Evaluation metrics further validated these trends. The \textbf{Fréchet Inception Distance (FID)} and \textbf{Kernel Inception Distance (KID)} steadily improved with training for all models. While the classical GAN attained the lowest FID score, the HQCGAN with 7 qubits achieved comparable results in later epochs.

Efficiency analysis showed that the usage of quantum shots remained fixed according to the HQCGAN model, regardless of the complexity of the data. Training time scaled linearly with the number of qubits, with HQCGAN-3 offering the fastest runtime amongst the HQCGANs. In general, the findings generally support the hypothesis that HQCGANs offer a viable generative alternative within the constraints of the \textbf{Noisy Intermediate-Scale Quantum (NISQ)} capabilities, balancing quantum expressiveness with classical efficiency.

\section{Discussion}

\subsection{Interpretation of Results}

The experimental results in Section~\ref{Experimental Results and Analysis} demonstrate both expected and intriguing outcomes with respect to the performance of classical GANs and HQCGANs. The classical GAN consistently achieved the lowest FID and KID, validating its maturity as a generative framework for image synthesis compared to the HQCGANs. However, HQCGANs with higher qubit counts, particularly the 7-qubit variant, exhibited convergence trends that resembled classical performance, especially in terms of FID. This suggests that increasing quantum latent dimensionality leads to improved sample fidelity, likely because of the expanded expressiveness of the latent space due to higher qubit counts.

The training dynamics reinforces this interpretation. Although the 3-qubit HQCGAN showed unstable and higher discriminator losses, the 5-qubit and 7-qubit variants achieved more stable convergence, mirroring the smoother loss curves seen in the classical GAN. These patterns imply that hybrid quantum-classical architectures are not only viable but capable of learning meaningful data distributions, in the scenario that the quantum latent representation is sufficiently rich.

It is also worth noting the computational cost associated with quantum sampling. Despite promising qualitative results, the quantum sampling process introduces overhead due to shot-based sampling and circuit noise. However, given that all HQCGANs used a consistent shot count of 23,808 throughout training, the comparison remains fair and reveals the trade-offs between performance and quantum resource usage. The ability of the 7-qubit HQCGAN to reach near-classical levels with limited quantum dimensionality supports the theoretical claim that quantum systems can encode richer priors even with a relatively low parameter count.

In summary, the findings substantiate the theoretical proposition that hybrid quantum-classical architectures serve as a viable paradigm to bridge the computational divide between classical and quantum systems, effectively leveraging the complementary strengths of both frameworks. The fact that the HQCGANs improve with additional qubits highlights the latent dimensional scalability of quantum circuits, underscoring their potential to serve as powerful generative priors in adversarial settings.

\subsection{Implications for Quantum Machine Learning (QML)}

The results of this study offer several key insights into the rapidly evolving landscape of QML, particularly in the context of hybrid generative models. First, the viability of HQCGANs as a substitute or a supplement to classical latent priors underscores the relevance of quantum circuits in practical learning scenarios. Unlike QGANs that often require full quantum processing pipelines, hybrid approaches like HQCGANs allow classical discriminators to interpret quantum-sampled latent vectors, thereby lowering the barrier to near-term implementation on NISQ hardware.

Importantly, the observed performance gains with increased qubit dimensionality suggest that even small-scale quantum circuits can introduce beneficial stochasticity and richer priors into generative operations. This aligns with current QML theory that quantum systems may offer exponential representational advantages in encoding high-dimensional probability distributions, especially when classical models are bottlenecked by latent space limitations.

Furthermore, the ability to train such models using backpropagation-compatible TensorFlow-based workflows illustrates a practical pathway for integrating quantum modules into mainstream machine learning pipelines. This opens the door to broader adoption and integration of QML, where hybrid systems could be used to augment classical pipelines in domains such as anomaly detection, data augmentation, and privacy-preserving generative modelling.

In summary, the findings reinforce the notion that hybrid QML systems are not merely theoretical constructs but are applicable in practical, data-driven contexts. As quantum hardware continues to evolve, architectures such as HQCGANs may represent a stepping stone towards the realisation of quantum-accelerated generative AI.

\subsection{Limitations of the Study}\label{limitations}

Despite promising outcomes, this study is subject to several limitations. Firstly, all quantum components were executed using the Qiskit AerSimulator rather than actual quantum hardware. Although this enabled precise control over noise and reproducibility, it fails to capture the full stochastic behaviour and decoherence present in real quantum devices.

Secondly, the study was constrained to a relatively low qubit count (3, 5, and 7 qubits) due to the limitations of the classical simulation resource. This restricts the expressive power of the latent space of the quantum generator, and it remains uncertain how the performance of HQCGAN would scale with higher qubit counts on actual quantum systems.

Furthermore, the experiments were limited to standard GAN architectures. Advanced variants such as Wasserstein GAN (WGAN), Deep Convolutional GAN (DCGAN), or StyleGAN were not employed, thus constraining our ability to benchmark the true relative advantage of HQCGANs over state-of-the-art classical models.

Lastly, both FID and KID rely on features extracted from the InceptionV3 network, which may not be optimal for evaluating synthetic MNIST-style images. This introduces potential bias in the evaluation metrics and may not fully reflect the perceptual quality.

\section{Conclusion and Future Work}

\subsection{Conclusion}

This research explored the viability and performance of Hybrid Quantum-Classical Generative Adversarial Networks (HQCGANs) in the context of generative modelling, using a filtered binary version of the MNIST dataset (digits 0 and 1). By integrating quantum-generated latent vectors with classical discriminators, the study sought to investigate whether quantum-enhanced sampling could offer advantages over purely classical GANs.

Classical GANs served as a performance baseline, and HQCGANs were trained with varying qubit counts (3, 5, and 7), each implemented using Qiskit’s AerSimulator with realistic noise modelling. All models shared similar network architectures, allowing fair comparisons. Performance was evaluated using standard generative metrics: Fréchet Inception Distance (FID), Kernel Inception Distance (KID), and Training Efficiency.

The results showed that HQCGANs with 5 and 7 qubits achieved competitive performance relative to the classical baseline, particularly in terms of FID and KID in later training epochs. Notably, the 7-qubit HQCGAN demonstrated near-classical performance, supporting the hypothesis that quantum-enhanced latent representations may improve generative modelling when scaled appropriately. Although the 3-qubit HQCGAN showed limitations due to reduced expressiveness, all HQCGANs remained stable throughout training.

This study offers a proof-of-concept for hybrid quantum-classical generative learning. Although classical models remain dominant under current conditions, the findings highlight the potential of quantum-assisted models, especially as quantum hardware matures. The implications are particularly relevant for tasks where high-dimensional data representations are expensive to learn classically, suggesting that quantum-enhanced sampling could serve as a valuable component in future generative learning pipelines.

\subsection{Future Research Directions}

Future research can address the limitations stated in Section~\ref{limitations} and further enhance the validity and impact of HQCGAN studies. A critical next step is to deploy HQCGANs on real quantum hardware, such as IBM Quantum systems or AWS Braket, to evaluate performance under realistic quantum noise and decoherence conditions. This would also allow experimentation with quantum error mitigation techniques to improve reliability.

Moreover, expanding the hybrid framework to incorporate quantum-based discriminators could enable the exploration of fully quantum GAN architectures and their training dynamics. Such an approach could provide deeper insights into the role of quantum entanglement in adversarial learning.

To strengthen empirical comparisons, future work should include state-of-the-art classical GAN variants such as WGAN-GP or StyleGAN. These benchmarks would more rigorously test whether quantum components offer tangible advantages over powerful classical architectures.

Finally, incorporating more sophisticated quantum noise models or even real-time error mitigation techniques would bring the experiments closer to practical deployment and deepen understanding of how HQCGANs function under noisy quantum conditions.

\section*{Acknowledgment}

I express my deepest gratitude to my lecturer, Mr. Gerald Chua, whose steadfast mentorship and unwavering support were instrumental throughout the development of this paper. From the early stages of topic formulation to the final phases of refinement, Mr. Chua consistently provided clear academic direction, constructive criticism, and invaluable encouragement. His ability to challenge my thinking while remaining patient and approachable greatly enriched both the theoretical and experimental rigour of this research. Beyond his role as a lecturer, his genuine dedication to the growth of his students has left a lasting impression on me. For his countless hours of guidance, often extending beyond regular academic duties, I am sincerely thankful.

I am also grateful to the School of Computing, Singapore Polytechnic, for equipping me with the infrastructure, tools, and resources needed to carry out this project. The availability of GPU-accelerated environments through AI and Analytical Colab significantly enabled the execution of complex experiments and deep learning workflows.

To my friends, thank you for offering your encouragement and keeping me grounded during this journey. Whether through meaningful conversations, moral support, or shared silence during moments of stress, your presence helped me persevere through the more demanding phases of this research.

I would especially like to acknowledge my parents for their unwavering belief in me and for supporting the financial costs of my cloud computing experiments. Their quiet sacrifices and continued encouragement made it possible for me to push the boundaries of this project further than I initially envisioned.

Lastly, I give thanks to God, whose providence, grace, and sustaining presence accompanied me at every step of this process. This paper stands as a humble offering and a testament to His faithfulness. \textit{Soli Deo Gloria}.

\end{document}